\documentclass[sigconf]{acmart}

\AtBeginDocument{%
  \providecommand\BibTeX{{%
    \normalfont B\kern-0.5em{\scshape i\kern-0.25em b}\kern-0.8em\TeX}}}

\usepackage{array}
\usepackage{subfigure}
\usepackage{graphicx}
\usepackage[switch]{lineno}
\usepackage{amsmath}
\usepackage{amsthm}
\usepackage{multirow}
\usepackage{booktabs}
\usepackage{bbding}
\usepackage{algorithm}
\usepackage{algorithmic}
\usepackage{xcolor}
\usepackage{xspace}
\usepackage{arydshln}
\usepackage[inline]{enumitem}
\usepackage{multicol}
\newcommand{\paratitle}[1]{\vspace{1.5ex}\noindent\textbf{#1}}
\newcommand{\ie}{\emph{i.e.,}\xspace}

\newcommand{\eg}{\emph{e.g.,}\xspace}
\newcommand{\baby}{{DIMO}\xspace}
\newcommand{\babyx}{{DIMO}}

\newcommand{\fig}{Figure\xspace}
\newtheorem*{template1}{Co-occurrence Template}
\newtheorem*{template2}{Feature Template}

\copyrightyear{2024}
\acmYear{2024}
\setcopyright{acmlicensed}\acmConference[SIGIR '24]{Proceedings of the 47th International ACM SIGIR Conference on Research and Development in Information Retrieval}{July 14--18, 2024}{Washington, DC, USA}
\acmBooktitle{Proceedings of the 47th International ACM SIGIR Conference on Research and Development in Information Retrieval (SIGIR '24), July 14--18, 2024, Washington, DC, USA}
\acmDOI{10.1145/3626772.3657748}
\acmISBN{979-8-4007-0431-4/24/07}

\begin{document}

\title[Disentangling ID and Modality Effects for Session-based Recommendation]{Disentangling ID and Modality Effects for\\ Session-based Recommendation}

\author{Xiaokun Zhang}
\affiliation{%
  \institution{Dalian University of Technology}
  \city{}
  \country{}
}
\email{dawnkun1993@gmail.com}

\author{Bo Xu}
\affiliation{%
  \institution{Dalian University of Technology}
  \city{}
  \country{}
  }
\email{xubo@dlut.edu.cn}

\author{Zhaochun Ren}
\affiliation{%
  \institution{Leiden University}
  \city{}
  \country{}
  }
\email{z.ren@liacs.leidenuniv.nl}

\author{Xiaochen Wang}
\affiliation{%
  \institution{Pennsylvania State University}
    \city{}
  \country{}
  }
\email{xcwang@psu.edu}

\author{Hongfei Lin}
\authornote{Corresponding Author.}
\affiliation{%
  \institution{Dalian University of Technology}
  \city{}
  \country{}
  }
\email{hflin@dlut.edu.cn}

\author{Fenglong Ma}
\affiliation{%
  \institution{Pennsylvania State University}
  \city{}
  \country{}
  }
\email{fenglong@psu.edu}

\renewcommand{\shortauthors}{Xiaokun Zhang, et al.}

\begin{abstract}
Session-based recommendation aims to predict intents of anonymous users based on their limited behaviors. 
Modeling user behaviors involves two distinct rationales: co-occurrence patterns reflected by item IDs, and fine-grained preferences represented by item modalities (\eg text and images).  
However, existing methods typically entangle these causes, leading to their failure in achieving accurate and explainable recommendations. 
To this end, we propose a novel framework \baby to disentangle the effects of ID and modality in the task. 
At the item level, we introduce a co-occurrence representation schema to explicitly incorporate co-occurrence patterns into ID representations.
Simultaneously, \baby aligns different modalities into a unified semantic space to represent them uniformly. 
At the session level, we present a multi-view self-supervised disentanglement, including proxy mechanism and counterfactual inference, to disentangle ID and modality effects without supervised signals. 
Leveraging these disentangled causes, \baby provides recommendations via causal inference and further creates two templates for generating explanations.
Extensive experiments on multiple real-world datasets demonstrate the consistent superiority of \baby over existing methods. Further analysis also confirms \baby's effectiveness in generating explanations.

\end{abstract}



\maketitle

\section{Introduction}

\begin{figure}[!t]
  \centering
  \includegraphics[width=0.93\linewidth]{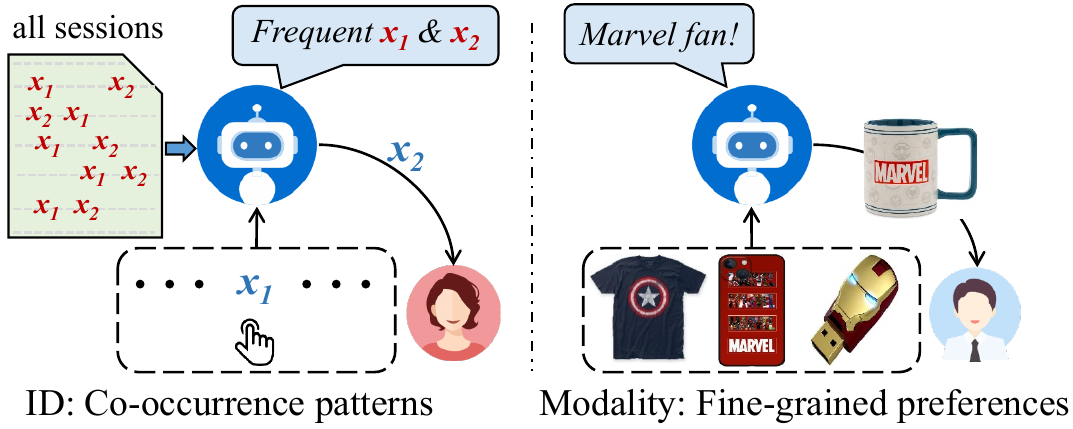}
  \caption{Two distinct rationales for modeling user behaviors: co-occurrence patterns of ID; fine-grained preferences of modality.}\label{example}
\vspace{-0.1in}
\end{figure}

Session-based recommendation (SBR) is designed for discerning intents of anonymous users through their limited behaviors within a brief period of time~\cite{GRU4Rec, NARM}. 
Conventional recommendations typically assume the availability of user profiles, enabling them to offer services based on users' long-term behaviors. However, acquiring user identities has become challenging due to widespread unlogged-in cases and stringent privacy regulations. Given its substantial practical potential, thus, SBR has recently garnered significant attention in both academia and industry~\cite{STAMP, Song@WWW2023, Yang@SIGIR2023}.

Deep neural networks, owing to their robust capabilities in representing items and sessions, are prevailing in SBR~\cite{Zhang@WSDM2023, Wang@CS2022}. Early efforts primarily focus on employing various networks to uncover co-occurrence patterns based on item \textbf{ID}, such as Recurrent Neural Networks (RNN)~\cite{GRU4Rec, NARM}, attention mechanism~\cite{STAMP, BERT4Rec} and Graph Neural Networks (GNN)~\cite{SR-GNN, LESSR}. 
More recently, a parallel line of research has emerged, aiming to incorporate item \textbf{modality} to enhance user intent learning including description text~\cite{hou@KDD2022}, images~\cite{Pan@CIKM2022, MMSBR, MMMLP}, and categories~\cite{Chen@SIGIR2023, cai@SIGIR2021}. The incorporation for modality information brings substantial improvements for SBR since it enables methods to capture user preferences form multiple perspectives and alleviate sparsity issue inherent in the task~\cite{CoHHN, Yang@SIGIR2023}. However, our investigation reveals a fundamental distinction between ID and modality on modeling user behaviors.

In fact, item \textbf{ID} merely acts as a symbolic identifier, failing to capture detailed item features~\cite{MMSBR, Yuan@SIGIR2023}. It reflects item \textbf{co-occurrence patterns} across all users' interactions from a statistical perspective. That is, corresponding methods rely on \emph{the wisdom of crowd} for predictions. 
As illustrated in the left part of \fig~\ref{example}, co-occurrence patterns of ID indicate a underlying principle in recommendation: if items $x_1$ and $x_2$ are frequently co-purchased, it is a sensible strategy that recommending $x_2$ to a user after she engaged with $x_1$.

On the contrary, item \textbf{modality}, \eg narrative text and images, can characterize item features such as style and color of clothes~\cite{Pan@CIKM2022, MMMLP, zhang2024side}. It represents user \textbf{fine-grained preferences} and empowers models to offer \emph{customized services tailored to individual interest}. As depicted in the right part of \fig~\ref{example}, analyzing images of purchased items might reveal that the user is a Marvel fan. Naturally, it would be prudent to recommend a co-branded Marvel cup, even if T-shirt and cup are rarely bought together. This underscores another rationale behind recommendation: crucial insights into user fine-grained preferences can be derived from item modality, guiding the presentation of items with similar attributes accordingly.

ID and modality undoubtedly encapsulate distinct rationales in unveiling user intents. However, existing methods entangle these causes in recommendation, frequently merging ID and modality embeddings indiscriminately~\cite{lai@SIGIR2022, MMSBR, MMMLP}. Thus, they fail to grasp underlying reasons guiding user actions, leading to inferior accuracy and interpretability. Despite being imperative, disentangling ID and modality is challenging due to following obstacles:

(1) \emph{How to represent ID and modality at item level effectively?} Current methods employ various neural structures, \eg RNN~\cite{NARM}, attention~\cite{BERT4Rec} and GNN~\cite{SR-GNN}, to encode co-occurrence patterns for item ID. Nevertheless, constrained by implicit learning paradigm, these methods struggle to adequately integrate co-occurrence patterns into ID embeddings. In addition, there exists a semantic gap between distinct modalities~\cite{Pan@CIKM2022, MMMLP}, making it hard to align these modalities and represent them uniformly.

(2) \emph{How to disentangle ID and modality effects at session level without supervised signals?} A user's selection is a latent variable reflecting either co-occurrence patterns of ID or fine-grained preferences of modality. However, there are no explicit signals that indicate which factor dominates user choice within a session. 
Hence, the absence of supervised signals poses a challenge in distinguishing the effects of ID and modality on user decisions. 

(3) \emph{How to improve SBR on both accuracy and interpretability?} 
On one hand, improving SBR accuracy based on distinct causes remains a daunting problem.
On the other hand, while interpretability holds significant importance in recommendation~\cite{Geng@WWW2022}, limited efforts have been dedicated to explainable SBR. Existing attempts rely on well-designed knowledge graphs~\cite{Wu@ICDE2023, Zheng@SIGIR2022} to generate explanations. However, these explanations, lacking understanding for rationales behind user actions, tend to remain superficial and unconvincing. Thus, implementing rationales derived from ID and modality for explainable SBR is also an urgent issue to be solved.

To tackle these issues, we propose a novel framework \baby, aiming to \underline{D}isentangle \underline{I}D and \underline{MO}dality effects to improve both accuracy and interpretability in session-based recommendation. 
At item level, we devise a \emph{co-occurrence representation schema} that explicitly incorporates co-occurrence patterns into ID embeddings. It constructs a \emph{global co-occurrence graph} to encode co-occurrence patterns, and presents a \emph{co-occurrence constraint} to highlight the patterns in ID representation learning.
Meanwhile, we convert different modalities into the same semantic space via techniques of Natural Language Processing (NLP) and Computer Vision (CV), enabling a unified modality representation. 
For session level, we present a \emph{multi-view self-supervised disentanglement} to distinguish ID and modality effects in the absence of supervised signals. It comprises two components: \emph{proxy mechanism}, where co-occurrence and modality proxies are established to guide cause learning; and \emph{counterfactual inference}, where we capture the effects of a single cause by eliminating the influence of the other. Subsequently, both causes are employed for predictions through \emph{causal inference}. Additionally, two templates, \ie \emph{co-occurrence template} and \emph{feature template}, are created to generate explanations. In summary, our contributions are as follows:

\begin{itemize}
    \item We identify and disentangle ID and modality effects on modeling user behaviors. To our best knowledge, this study marks the first attempt to disentangle co-occurrence patterns of ID and fine-grained preferences of modality in SBR.
    \item A novel \baby is proposed for ID and modality disentanglement, where several innovative techniques are presented to distinguish ID and modality at item and session levels, and improve both accuracy and interpretability of SBR.
    \item Extensive experiments over multiple public benchmarks demonstrate the superiority of our \baby over current state-of-the-art methods. Further analysis also justifies the effectiveness of \baby in generating meaningful explanations.
\end{itemize}

\section{Related Work}

\subsection{Session-based Recommendation}
\paratitle{ID-based methods} mine item co-occurrence patterns for recommendations. Most efforts employ various networks to capture the patterns like RNN~\cite{GRU4Rec, NARM}, attention mechanism~\cite{STAMP, BERT4Rec, DIDN, Zhou@CIKM2023}, GNN~\cite{SR-GNN, LESSR}, and MLP~\cite{zhou@WWW2022}. Another line of research further enhances co-occurrence learning via exploring extra sessions~\cite{pang@WSDM2022, Su@WWW2023,Qiao@CIKM2023}, multiple item relations~\cite{han@SIGIR2022, Zhang@WSDM2023, Li@WSDM2023}, multiple intents~\cite{guo@WSDM2022}, counterfactual augmentation~\cite{Song@WWW2023} and frequency domain~\cite{Du@SIGIR2023}. Unfortunately, only modeling item IDs, these methods fall short of recognizing user fine-grained preferences, limiting their performance.

\paratitle{Modality-enhanced methods} utilize modality information to reveal user intents. With rich semantics, description text is popular in characterizing item features and user preferences~\cite{hou@KDD2022, Li@KDD2023, Liu@CIKM2023Text}. Category, as an indicator for the user hierarchical decision-making process, is also widely explored in the task~\cite{lai@SIGIR2022, Zhou@CIKM2020, Chen@SIGIR2023, cai@SIGIR2021}. 
To capture user price sensitivity, some methods take item price into account~\cite{CoHHN, BiPNet}. 
Moreover, there are some pioneering efforts modeling images and text to finely-grained portray user interest~\cite{Pan@CIKM2022, MMMLP,Hu@CIKM2023, MMSBR}. However, entangling distinct rationales of ID and modality in capturing user intents, these methods fail to achieve optimal performance.

\subsection{Explainable Recommendation}
Interpretability of recommender systems is a hot-spot topic, as a reasonable explanation can greatly increase user satisfaction\cite{Geng@WWW2022, Zhang@SIGIR2014}. Nevertheless, explainable SBR is under-explored because it is hard to find reasons based on limited user behaviors. Recently, some pioneering efforts aim to explain SBR by introducing knowledge graphs (KG) where pathways are used to assist explanations\cite{Wu@ICDE2023, Chen@SIGIR2023}. Another work pre-defines handcraft rules, \eg repeat consumption, to provide explanations\cite{Chen@IJCNN2021}. However, all of them are laborious, like well-designed KG or selective rules, limiting their implementation. More importantly, failing to distinguish rationales of user actions, these methods fail to generate convincing explanations.

\subsection{Disentanglement in Recommendation}
Disentanglement learning tries to improve both model performance and interpretability by disentangling latent causes~\cite{Ma@NIPS2019}. Recent approaches conduct disentanglement on various causes in the context of recommendations such as user interest and conformity~\cite{Zheng@WWW2021}, long- and short-term interest~\cite{Zheng@WWW2022}, category-independent and -dependent preferences~\cite{Zhang@WSDM2023Disen}, as well as distinct layers of Transformer~\cite{Zhang@KDD2023}. However, none of the existing methods makes an effort to distinguish ID and modality effects on modeling user behaviors. Hence, the proposed \baby is the first to fill this research gap.


\section{Preliminaries}

\subsection{Problem Formulation}
Let $\mathcal{I}$ denote the item set, where $n = |\mathcal{I}|$ is the total number of items. For an item $x_i \in \mathcal{I}$ ($1 \leqslant i \leqslant n $), it can be represented from two views: ID, an integer indicating item co-occurrence associations, and modality that contains rich semantics, \ie $x_i$ = \{$x_i^{id}$, $x_i^{mo}$\}. In this work, we consider two common modalities including textual $x_i^{txt}$ and visual $x_i^{img}$ modality, \ie $x_i^{mo}$ = \{$x_i^{txt}$, $x_i^{img}$\}, where $x_i^{txt}$ may contain title and brand, and $x_i^{img}$ is item image. The $x_i^{txt}$ consists of a sequence of words as $x_i^{txt}$ = \{$w_1$, $w_2$, ..., $w_t$\}. A session $\mathcal{S}$ = [$x_1, x_2, ..., x_m$] is generated by an anonymous user within a certain period of time, where $x_i$ $\in$ $\mathcal{I}$ and $m$ is the session length. The goal of \baby is to predict next interacted item $x_{m+1}$ based on $\mathcal{S}$ while generating reasonable explanations.

\subsection{Global Co-occurrence Graph Construction}

\begin{figure}[t]
  \centering
  \includegraphics[width=0.95\linewidth]{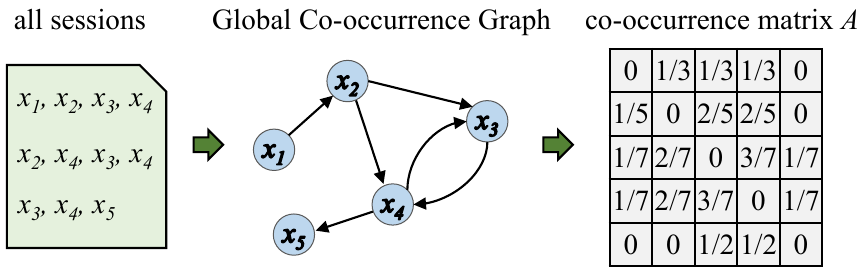}
  \caption{Global co-occurrence graph construction.}\label{gcg}
\end{figure}

GNN is effective on item representation learning in SBR~\cite{SR-GNN,GCE-GNN,Andreas@RecSys2023}. 
Thus, we further construct a \emph{global co-occurrence graph} to record all item co-occurrence associations in session data. Specifically, the global co-occurrence graph is a weighted directed graph $\mathcal{G}$ = $(V, A)$, where $V = \mathcal{I}$ is node set, and $A \in \mathbb{R}^{n \times n}$ is co-occurrence matrix. As illustrated in~\fig~\ref{gcg}, the weight $a_{ik}$ of $A$ between item $x_i$ and $x_k$ is defined based on their co-occurrence frequency via,
\begin{align}
    a_{ik} &= \frac{count(x_i, x_k)}{\sum_{x_j \in N_{i}}count(x_i, x_j)},
\end{align}
where $count(x_i, x_k)$ counts the number of times two items co-occurred in all sessions, and $N_i$ is an item set consisting of items that have appeared in a session with $x_i$. Based on the co-occurrence matrix $A$, we can easily identify co-occurrence associations between items. For instance, the $j$-th row of $A$ represents the co-occurrence associations of item $x_j$ with all other items, where the larger the value, the higher their co-occurrence frequency.

\section{The Proposed \baby}

\begin{figure*}[t]
  \centering
  \includegraphics[width=0.93\linewidth]{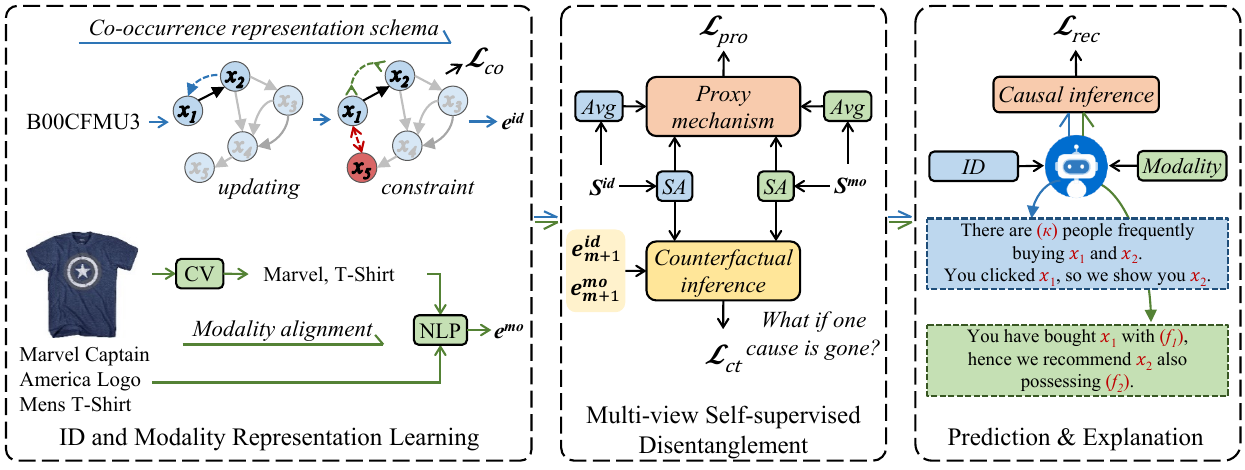}
  \caption{The architecture of \baby. ID and modality representation learning explicitly incorporates co-occurrence patterns into ID embeddings while conducting modality alignment for unified modality representation. Multi-view self-supervised disentanglement distinguishes ID and modality effects via proxy mechanism and counterfactual inference. Based on disentangled causes, \baby provides recommendation via causal inference and generates explanations on two kinds of templates.}\label{DIMO}
\end{figure*}

As illustrated in \fig~\ref{DIMO}, the proposed \baby mainly consists of the following components: (1) ID and modality representation learning aims to effectively represent item ID and modality; (2) multi-view self-supervised disentanglement distinguishes ID and modality effects on modeling user behaviors; (3) prediction forecasts items of interest via causal inference; and (4) explanation generates explanations based on disentangled causes.

\subsection{ID and Modality Representation Learning}
In this part, we aim to explicitly incorporate co-occurrence patterns into ID embeddings and align distinct modalities for unified modality representation.

\subsubsection{Co-occurrence Representation Schema}
Following common paradigm~\cite{NARM, SR-GNN}, we build an embedding table $\mathbf{\hat{E}}^{id} \in \mathbb{R}^{n \times d}$ to store ID semantics. To encode co-occurrence patterns, we first employ the \emph{global co-occurrence graph} for ID embedding updating via,
\begin{align}
    \mathbf{E}^{id} &= (A+I)\mathbf{\hat{E}}^{id},
\end{align}
where $A$ is a co-occurrence matrix and $I \in \mathbb{R}^{n \times n}$ is an identity matrix. For an item ID, we repeat Eq.(2) $c$ times to update its embedding based on all items co-occurring with it, where the importance is determined by their co-occurrence frequency. Afterwards, for an item ID $x_i^{id}$, we can obtain its ID embedding $\mathbf{e}_i^{id} \in \mathbb{R}^{d}$ based on $\mathbf{E}^{id}$. 

It is intuitive that item IDs with high co-occurrence frequency tend to have similar semantics. Thus, a \emph{co-occurrence constraint} is presented to drive frequently co-occurring IDs close while rarely co-occurring ones apart. For an item ID $\mathbf{e}_i^{id}$, we build a positive set $\{\mathbf{e}_1^{id+}, ..., \mathbf{e}_l^{id+}\}$ containing item IDs with top-$l$ co-occurrence frequency with the item based on $A$. The negative set $\{\mathbf{e}_1^{id-}, ..., \mathbf{e}_l^{id-}\}$ contains item IDs not co-appearing with $\mathbf{e}_i^{id}$ in any sessions. Formally, we conduct co-occurrence constraint for item ID as,
\begin{align}
    \mathcal{L}_{co} &= -\frac{{\rm sim}(\mathbf{e}_i^{id}, \mathbf{\bar{e}}_i^+)}{\sum_{k=1}^{l}{\rm sim}(\mathbf{e}_i^{id}, \mathbf{e}_k^{id-})}, 
\end{align}
where ${\rm sim}(\cdot)$ is cosine similarity and $\mathbf{\bar{e}}_i^+ = \frac{1}{l}\sum_{k=1}^l \mathbf{e}_k^{id+}$. Note that, Eq.(3) brings embeddings of frequently co-occurring IDs close while separating rarely co-occurring ones. Such a schema ensures that an item ID embedding tends to be similar with its frequent co-occurrence items instead of rare co-occurrence ones. Thus, it highlights co-occurrence patterns in ID representation learning.

\subsubsection{Modality Alignment}
There exists a huge semantic gap between item text and images~\cite{MMSBR, MMMLP}. In \baby, we convert item images into text for modality alignment. We align distinct modalities into text space due to: (1) text has been proven effective in representing user preferences~\cite{hou@KDD2022, Li@KDD2023}; (2) the model can benefit from the development of NLP to effectively represent text; and (3) text is easy to grasp and contributes to generating clear explanations. Concretely, for an item image $x_i^{img}$, we utilize GoogLeNet to conduct image classification with 1,000 categories on it, and the predicted top-2 categories are cascaded as its generated text $x_i^{gtxt}$ = \{$w_1^{\prime}$, ..., $w_r^{\prime}$\}. Then, we joint $x_i^{txt}$ and $x_i^{gtxt}$ to formulate $x_i^{mo}$ = \{$w_1$, ..., $w_t$, $w_1^{\prime}$, ..., $w_r^{\prime}$\}. BERT~\cite{BERT} is then employed on every word in $x_i^{mo}$ to obtain word embedding sequence \{$\mathbf{e}_1$, ..., $\mathbf{e}_{t+r}$\}. Finally, an item modality information is represented by,
\begin{align}
    \mathbf{e}_i^{mo} &= \frac{1}{t+r}\sum_{k=1}^{t+r} \mathbf{e}_k ,
\end{align}
where $\mathbf{e}_i^{mo} \in \mathbb{R}^{d}$ is the unified embedding for modalities.

\subsubsection{Sequence Encoding} Based on the learned ID embedding $\mathbf{e}_i^{id}$ and modality embedding $\mathbf{e}_i^{mo}$, we can formulate a session from two views: $\mathcal{S}^{id}$ = [$\mathbf{e}_1^{id}, ..., \mathbf{e}_m^{id}$], and $\mathcal{S}^{mo}$ = [$\mathbf{e}_1^{mo}, ..., \mathbf{e}_m^{mo}$]. Self-attention is effective in representing user behavior sequences~\cite{BERT4Rec, Zhou@CIKM2023}, so we employ it to handle these sequences as follows,
\begin{align}
    \mathbf{s}^{id} &= SA_1(\mathcal{S}^{id}), \\
    \mathbf{s}^{mo} &= SA_2(\mathcal{S}^{mo}),
\end{align}
where $SA_1$ and $SA_2$ signify self-attention layers with different parameters, and $\mathbf{s}^{id}$, $\mathbf{s}^{mo} \in \mathbb{R}^d$ represent co-occurrence patterns and user fine-grained preferences at the session level, respectively. 

\subsection{Multi-view Self-supervised Disentanglement}

In this part, we propose a multi-view self-supervised disentanglement, including proxy mechanism and counterfactual inference, to distinguish ID and modality effects at session level.

\subsubsection{Proxy Mechanism}
The proxy mechanism aims to find a proxy for each cause where the cause and its proxy possess similar semantics. After that, we can push the cause and its proxy close while pulling different causes apart for cause disentanglement. Specifically, we perform average-pooling on  $\mathcal{S}^{id}$ and $\mathcal{S}^{mo}$ to obtain proxies via,
\begin{align}
    \mathbf{\hat{s}}^{id} &= Avg([\mathbf{e}_1^{id}, \mathbf{e}_2^{id}, ..., \mathbf{e}_m^{id}]), \\
    \mathbf{\hat{s}}^{mo} &= Avg([\mathbf{e}_1^{mo}, \mathbf{e}_2^{mo}, ..., \mathbf{e}_m^{mo}]),
\end{align}
where $Avg(*)$ is the average operation, and $\mathbf{\hat{s}}^{id}$, $\mathbf{\hat{s}}^{mo} \in \mathbb{R}^d$ are proxies for co-occurrence patterns and user fine-grained preferences, respectively. Note that, we employ different methods to represent cause (self-attention) and its proxy (average-pooling). The reasons are twofold: (1) aggregating information within a session via different methods makes the learned embeddings exhibit subtle differences while retaining overall similar semantics, contributing to subsequent disentanglement; and (2) average operations are parameter-free without introducing additional complexity into \baby.
Formally, we disentangle different causes via,
\begin{equation}
    \begin{split}
    \mathcal{L}_{pro} &= -\frac{{\rm sim}(W_1\mathbf{s}^{id}, W_2\mathbf{\hat{s}}^{id})}{{\rm sim}(W_1\mathbf{s}^{id}, W_2\mathbf{\hat{s}}^{id}) + {\rm sim}(W_3\mathbf{s}^{id}, W_4\mathbf{s}^{mo})}\\
    &-\frac{{\rm sim}(W_5\mathbf{s}^{mo}, W_6\mathbf{\hat{s}}^{mo})}{{\rm sim}(W_5\mathbf{s}^{mo}, W_6\mathbf{\hat{s}}^{mo}) + {\rm sim}(W_7\mathbf{s}^{mo}, W_8\mathbf{s}^{id})},
    \end{split}
\end{equation}
where $W_* \in \mathbb{R}^{d \times d}$ are trainable parameters. As shown in Eq.(9), we urge one cause to tend to its proxy while away from the other cause. Such a mechanism enlarges the gap between different causes, thus achieving cause disentanglement.

\subsubsection{Counterfactual Inference} Counterfactual inference is a process aimed at understanding hypothetical outcomes under alternative actions, contrary to what was observed under specific conditions. This process contributes to clarifying causal relations. In this study, we pose a counterfactual question to aid in cause disentanglement: "What if ID or modality cause is gone?" If one cause is eliminated, we can infer that a user action is determined by the remaining cause. 
Thus, signals are pinpointed for cause disentanglement.
Besides, item co-occurrence associations can be easily examined based on the co-occurrence matrix $A$. In light of these, we rely on $A$ for conducting counterfactual inference. Specifically, for an item $x_i^{id}$, we can obtain its co-occurrence set $N_i$ consisting of items whose value is nonzero in the $i$-th row of $A$. In a session [$x_1^{id}, ..., x_m^{id}$], co-occurrence sets of all items are joined to obtain $N_s$ = $\bigcup N_i$. We then distinguish ID and modality effects via,
\begin{align}
\mathcal{L}_{ct}=
\begin{cases}
    -\frac{{\rm sim}(\mathbf{s}^{mo}, \mathbf{e}_{m+1}^{mo})}{{\rm sim}(\mathbf{s}^{mo}, \mathbf{e}_{m+1}^{mo}) + {\rm sim}(\mathbf{s}^{id}, \mathbf{e}_{m+1}^{id})}, & x_{m+1} \notin N_s
    \\
    -\frac{{\rm sim}(\mathbf{s}^{id}, \mathbf{e}_{m+1}^{id})}{{\rm sim}(\mathbf{s}^{id}, \mathbf{e}_{m+1}^{id}) + {\rm sim}(\mathbf{s}^{mo}, \mathbf{e}_{m+1}^{mo})}, & x_{m+1} \in N_s
\end{cases}
\end{align}
where $x_{m+1}$ is label item, and $\mathbf{e}_{m+1}^{id}$, $\mathbf{e}_{m+1}^{mo} \in \mathbb{R}^d$ denote its ID and modality embeddings, respectively. As presented in Eq.(10), if the label item has no co-occurrence associations with items in the current session, we can infer that the user selects it due to her fine-grained preferences behind modalities. That is, modality effects dominate user choice. Hence, the similarity of $\mathbf{s}^{mo}$ and $\mathbf{e}_{m+1}^{mo}$ should be larger than that of $\mathbf{s}^{id}$ and $\mathbf{e}_{m+1}^{id}$. Besides, if $x_{m+1} \in N_s$, we simply attribute user decisions as co-occurrence patterns because co-occurrence behaviors are common and easy to explain. 
Such a schema identifies the dominating cause in influencing user choice, thus facilitating ID and modality disentanglement.

\subsection{Prediction with Causal Inference}
As stated before, the next actions of a user mainly originate from two causes: co-occurrence patterns of ID or fine-grained preferences of modality. 
With the ability to deduce results from causes, causal inference~\cite{Zheng@WWW2021, Zheng@WWW2022} is used here to predict items of interest for users. 
Formally, for a session ($\mathbf{s}^{id}$, $\mathbf{s}^{mo}$) and a candidate item ($\mathbf{e}_i^{id}$, $\mathbf{e}_i^{mo}$), their interaction probability can be predicted by,
\begin{align}
    y_i &= \mathbf{s}^{id}  \mathbf{e}_i^{id} + \mathbf{s}^{mo}  \mathbf{e}_i^{mo},
\end{align}
where we instantiate causal inference with addition operation due to: (1) it is straightforward; and (2) it follows the logic of recommendation, \ie an item should be shown as long as it reflects item co-occurrence patterns or user fine-grained preferences. Then, the recommendation task is optimized by cross-entropy as,
\begin{align}
    \mathcal{L}_{rec} = - \sum^n_{i=1} p_i \log (y_i) + (1-p_i)\log(1-y_i),
\end{align}
where $p_i$ is ground truth signifying whether a user engages with item $x_i$. Finally, \baby is trained by multi-task learning schema as,
\begin{align}
    \mathcal{L} = \mathcal{L}_{rec} + \lambda (\mathcal{L}_{co} + \mathcal{L}_{pro} + \mathcal{L}_{ct}),
\end{align}
where $\lambda$ balances the recommendation task and auxiliary tasks.

\subsection{Explanation}
Based on disentangled causes, we build explainable SBR by creating a co-occurrence template and a feature template.
\subsubsection{Co-occurrence template} 
Co-occurrence association, as a principle for recommendation, is first employed to generate explanations. 
Based on recommended item $x_{m+1}$, an item $x_i$ is selected from current session $S$ if $count(x_{m+1}, x_i)$ is larger than a constant $\eta$. After that, we can form the co-occurrence template as, 
\begin{template1}
    There are ($\kappa$) people frequently buying $x_i$ and $x_{m+1}$ together. You clicked $x_i$, so we show you $x_{m+1}$.
\end{template1}
\noindent where $\kappa$=$count(x_{m+1}, x_i)$. Note that, $count(x_{m+1}, x_i)$ is determined based on all sessions, indicating that co-occurrence patterns represent the wisdom of crowd. 

\subsubsection{Feature template}
Suggesting another rationale of recommendation, modality presents user fine-grained preferences on certain item features. Thus, we build another explanation template at the feature level. 
Concretely, based on aligned modalities, text features are available for an item $x_i$ as $x_i^{mo}$ = \{$\mathbf{e}^{i}_1$, ..., $\mathbf{e}^{i}_{t+r}$\}. For a session $\mathcal{S}$ = [$x_1, ..., x_m$] and its recommended item $x_{m+1}$, we calculate cosine similarity between each pair of words in $x_i^{mo} \in S$ and $x^{mo}_{m+1}$. The text pairs with similarity larger than $\gamma$ are viewed as features of interest. 
Formally, we build the feature template as,
\begin{template2}
    You have bought $x_i$ with ($f_1$), hence we recommend $x_{m+1}$ also possessing ($f_2$).
\end{template2}
\noindent where ($f_1$, $f_2$) is a pair of text features with similar semantics and $f_1 \in x_i^{mo}$, $ f_2 \in x_{m+1}^{mo}$. 
Note that, the time complexity of this schema is affordable because an item text (like title) usually contains a few words, and a session also consists of a few items. Besides, feature pairs are selected within items of the current session, which signifies that modality presents strong customized characteristics.


\subsubsection{Template selection}
Based on above two kinds of templates, we clarify our selection for template as follows,
\begin{footnotesize}
    \begin{align}
     template :=
    \begin{cases}
        co-occurrence  & \exists count(x_i, x_{m+1}) > \eta, x_i \in S \\
        feature  & \exists {\rm sim}(f_1, f_2) > \gamma, f_1 \in x_i^{mo}, f_2 \in x_{m+1}^{mo}
    \end{cases}
    \end{align}
\end{footnotesize}


\section{Experimental setup}



Extensive experiments are conducted to demonstrate the effectiveness of \baby via responding the following research questions:
\begin{enumerate*}[label=(\roman*)]
\item \textbf{RQ1}: How does \baby perform compared with existing state-of-the-art methods? (ref. Section~\ref{sec:overall})
\item \textbf{RQ2}: Does each proposed technique contribute positively to \baby's performance? (ref. Section~\ref{sec:emb}-\ref{sec:multidisen})
\item \textbf{RQ3}: Could \baby achieve disentanglement for ID and modality effects?  (ref. Section~\ref{sec:disen})
\item \textbf{RQ4}: How well are the explanations generated by \baby?  (ref. Section~\ref{sec:explanation})
\item \textbf{RQ5} What is the influence of hyper-parameters on \baby?  (ref. Section~\ref{sec:hyper})
\end{enumerate*}
\subsection{Datasets and Preprocessing}

\begin{table}[t]
\tabcolsep 0.05in 
\centering
\caption{Statistics of datasets.}
\begin{tabular}{ccccc}
\toprule
Datasets      & Cellphones & Grocery & Sports & Instacart \\
\midrule
\#item        & 9,091   & 7,286   & 14,650  & 10,009   \\
\#interaction & 123,186  & 151,251 & 282,102 &  380,230 \\
\#session     & 40,344  & 43,648   & 90,492   & 88,022 \\
avg.length    & 3.05     & 3.47     & 3.12    & 4.32 \\
\bottomrule
\end{tabular}

\label{statistics}
\end{table}

\begin{table*}[ht]
\tabcolsep 0.03in 
    \caption{Performance comparison among \baby and ten baselines over four datasets. The results (\%) produced by the best baseline and the best performer in each row are underlined and boldfaced, respectively. Significant improvements of \baby over the best baseline (*) is determined by the $t$-test ($p < 0.01$).}
    \renewcommand{\arraystretch}{1.1}
\begin{tabular}{cccccccccccccc}
\toprule 
Datasets                    & Metrics & SKNN  & NARM  & BERT4Rec    & SR-GNN & MSGIFSR     & Atten-Mixer    &MSGAT & MGS         & UniSRec     & MMSBR  & DIMO           & $impro.$   \\
\midrule 
\multirow{4}{*}{Cellphones} & Prec@10 & 14.31 & 15.42 & 18.31       & 16.36  & 17.80       & 19.51  & 17.22       & \underline{21.54} & 20.30       & 20.59  & $ \bf 31.66^*$ & 46.98\% \\
                            & MRR@10  & 8.84  & 12.43 & 11.96       & 12.96  & 12.40       & \underline{14.54}  & 13.41  & 14.24       & 14.32       & 13.94  & $ \bf 16.98^*$ & 16.78\% \\
                            & Prec@20 & 16.48 & 16.80 & 22.44       & 18.11  & 21.16       & 22.28  & 20.01       & \underline{25.02} & 23.78       & 22.82  & $ \bf 38.81^*$ & 55.12\% \\
                            & MRR@20  & 9.06  & 12.53 & 12.25       & 13.09  & 12.64       & \underline{14.71}  & 13.67  & 14.48       & 14.56       & 14.13  & $ \bf 17.36^*$ & 18.01\% \\
                            \midrule 
\multirow{4}{*}{Grocery}    & Prec@10 & 40.40 & 45.67 & 45.82       & 44.33  & 45.45       & 47.65  & 45.20       & 46.59       & \underline{47.95} & 46.05  & $ \bf 53.03^*$ & 10.59\% \\
                            & MRR@10  & 27.64 & 40.39 & 38.33       & 39.44  & 38.16       & 40.71  & 39.98      & 38.83       & \underline{40.90} & 39.01  & $ \bf 41.81^*$ & 2.22\%  \\
                            & Prec@20 & 42.40 & 47.14 & 49.07       & 46.24  & 48.15       & 49.56  & 47.01      & 48.37       & \underline{50.21} & 47.89  & $ \bf 57.01^*$ & 13.54\% \\
                            & MRR@20  & 27.78 & 40.59 & 38.56       & 39.64  & 38.35       & 40.84  & 40.12       & 38.98       & \underline{41.05} & 39.23  & $ \bf 41.98^*$ & 2.27\%  \\
                            \midrule 
\multirow{4}{*}{Sports}     & Prec@10 & 31.79 & 35.55 & {38.13} & 36.31  & 36.27       & 37.30  & 37.19       & 36.79       & \underline{38.31}       & 36.69  & $ \bf 45.07^*$ & 17.65\% \\
                            & MRR@10  & 24.23 & 33.40 & \underline{33.89} & 33.36  & 30.36       & 33.63    & 33.69       & 32.39       & 33.50       & 32.52  & $ \bf 34.86^*$ & 2.86\%  \\
                            & Prec@20 & 33.98 & 36.67 & {40.34} & 37.69  & 39.65       & 39.19      & 38.53     & 38.45       & \underline{40.62}       & 38.29  & $ \bf 49.86^*$ & 22.75\% \\
                            & MRR@20  & 24.39 & 33.57 & \underline{34.01} & 33.66  & 30.59       & 33.86    & 33.91      & 32.50       & 33.76       & 32.73  & $ \bf 35.15^*$ & 3.35\%  \\
                            \midrule 
\multirow{4}{*}{Instacart}  & Prec@10 & 6.78  & 8.27  & 10.92       & 8.96   & \underline{11.56} & 8.11     & 9.29       & 8.59        & 8.65        & 9.89   & $ \bf 12.51^*$ & 8.22\% \\
                            & MRR@10  & 2.06  & 3.02  & 3.58        & 3.27   & \underline{3.74}  & 3.12     & 3.54        & 2.87        & 3.21        & 3.61   & $ \bf 4.31^*$  & 15.24\% \\
                            & Prec@20 & 11.79 & 12.19 & 16.00       & 13.00  & \underline{16.44} & 11.53    & 13.36     & 13.74       & 12.55       & 14.37  & $ \bf 18.36^*$ & 11.68\% \\
                            & MRR@20  & 2.41  & 3.25  & 3.81        & 3.64   & \underline{4.02}  & 3.36     & 3.77        & 3.09        & 3.62        & 3.84   & $ \bf 4.81^*$ & 19.65\% \\
                            \bottomrule
\end{tabular}
\label{performance}
\end{table*}

    
    
    


We employ four public datasets to evaluate the performance of \baby and all baselines in SBR, including \textbf{Cellphones}, \textbf{Grocery}, \textbf{Sports}, and \textbf{Instacart}. The first three datasets cover different domains in Amazon\footnote{\url{http://jmcauley.ucsd.edu/data/amazon/}} and are popular in SBR~\cite{BERT4Rec, CoHHN}. As in~\cite{CoHHN}, a session is generated based on user behaviors that happened within one day. Instacart\footnote{\url{https://www.kaggle.com/c/instacart-market-basket-analysis}} is a competition dataset of Kaggle.
To simulate SBR scenarios, 20\% transactions with the least length are selected in Instascart. 
For Amazon datasets, the modality information includes text and images, and we use text for Instacart. 
For all datasets, the text contains title and brand. In a session, the last item is viewed as the label required to predict, and the remaining items are used to capture user intents. Following~\cite{NARM,SR-GNN}, sessions with length 1 and items appearing less than 5 times are filtered out. Each dataset is chronologically split into three parts for training, validation and testing with the ratio of 7:2:1. Statistical details of all datasets are shown in Table~\ref{statistics}.

\subsection{Evaluation Metrics}

Similar as~\cite{NARM, SR-GNN, lai@SIGIR2022, hou@KDD2022}, two common metrics, \textbf{Prec}@K (Precision) and \textbf{MRR}@K (Mean Reciprocal Rank), are utilized to evaluate the performance of \baby and all baselines, where K $\in$ \{10, 20\}. For both Prec@K and MRR@K, the larger the values, the better the performance of methods.




\subsection{Baselines}

We select following two groups of competitive baselines for performance comparison:

\paratitle{ID-based methods} focus on mining co-occurrence patterns of ID to provide recommendations: 
(1) \textbf{SKNN} determines next items based on item co-occurrence frequency in all sessions; 
(2) \textbf{NARM}~\cite{NARM} utilizes GRU with attention mechanism to capture users' main intents; 
(3) \textbf{BERT4Rec}~\cite{BERT4Rec} employs Transformer to model user sequential behaviors; 
(4) \textbf{SR-GNN}~\cite{SR-GNN} captures complex relations among items via GNN; 
(5) \textbf{MSGIFSR}~\cite{guo@WSDM2022} divides a session into multiple snippets to highlight co-occurrence patterns;
(6) \textbf{Atten-Mixer}~\cite{Zhang@WSDM2023} leverages multi-level user intents to achieve multi-level reasoning on item co-occurrence transitions.
(7) \textbf{MSGAT}~\cite{Qiao@CIKM2023} combines intra- and inter-session information to handle SBR.

\paratitle{Modality-enhanced methods} utilize modality information to capture user fine-grained preferences: 
(8) \textbf{MGS}~\cite{lai@SIGIR2022} exploits item categories for accurate preferences estimation; 
(9) \textbf{UniSRec}~\cite{hou@KDD2022} incorporates item text to obtain universal sequence representations;
(10) \textbf{MMSBR}~\cite{MMSBR} is a pioneering work incorporating both text and images to handle SBR.

\subsection{Implementation Details}
To ensure a fair comparison, we fix the embedding size of all methods at 100 ($d$=100). 
Other hyperparameters of \baby and all baselines are determined via grid search according to their performance on Prec@20 in validation set. 
For \baby, we search iterative times of global co-occurrence graph $c$ in $\{1, 2, 3, 4, 5\}$, and balance coefficient $\lambda$ in $\{0.1, 0.01, 0.001, 0.0001\}$.
Besides, the number of positives and negatives in the co-occurrence constraint is set as 10 ($l$=10), and the threshold in co-occurrence and feature template is 10 and 0.8, respectively (\ie $\kappa$=10, $\gamma$=0.8).
The mini-batch size is $50$, and learning rate is $0.001$. 
The source codes are available online\footnote{\url{https://github.com/Zhang-xiaokun/DIMO}}.

\section{Results and Analysis}
\subsection{Overall Performance (RQ1)}\label{sec:overall}

\begin{table}[t]
\tabcolsep 0.01in 
  \centering
    \caption{Effect of ID and modality representation learning.}
    \begin{tabular}{c cc cc cc}  
    \toprule  
    \multirow{2}*{Method}& 
    \multicolumn{2}{c}{Cellphones}&\multicolumn{2}{c}{Grocery}&\multicolumn{2}{c}{Sports}\cr  
    \cmidrule(lr){2-3} \cmidrule(lr){4-5} \cmidrule(lr){6-7} 
    &Prec@20&MRR@20 &Prec@20&MRR@20 &Prec@20&MRR@20\cr  
    \midrule  
    \babyx-co      &37.76&16.49   &55.88&41.23    &48.48&34.27\cr  
    \babyx$_{cas}$   &38.49&17.02   &56.63&41.64    &49.37&34.86\cr
	{\bf \baby} &{ $ \bf 38.81^*$ }&{$\bf 17.36^*$} 
	            &{$\bf 57.01^*$ }&{ $ \bf 41.98^*$ }
	            &{$\bf 49.86^*$}&{$\bf 35.15^*$ }
             \cr
	\bottomrule
    \end{tabular}

    \label{representation}
\end{table} 

We report the performance of \baby and all baselines in Table~\ref{performance}, where the following observations are noted:

Firstly, baseline methods present a significant performance gap across different contexts (\ie datasets). To name a few, BERT4Rec achieves competitive results in Sports, while performing poorly on Cellphones; and UniSRec surpasses all other baselines in Grocery, but its performance on Instacart is unsatisfactory. This phenomenon demonstrates how challenging it is to handle the task of SBR where methods are required to capture the intents of anonymous users based merely on their limited behaviors. It also proves the necessity and significance of studying this task. 

Secondly, among ID-based methods, recent methods, MSGIFSR and Atten-Mixer, achieve impressive performance. Both of these models highlight item co-occurrence associations by dividing a user behavior sequence into several sub-sequences. Such an operation enables them to capture multi-level co-occurrence patterns, contributing to their performance improvements. As to modality-enhanced methods, UniSRec generally performs well. We contend that the effectiveness of UniSRec benefits from its inclusion of informative item text, which enables it to gain some insights into user fine-grained preferences.

Thirdly, modality-enhanced methods obtain better accuracy over ID-based methods in most cases, \eg in Grocery. It is intuitive since that extra information provides opportunities for models to analyze user behaviors from various angles, obtaining much more insights on user behaviors. However, in some instances, like on the MRR metric in Cellphones and Sports, modality-enhanced methods are defeated by ID-based methods. We believe that existing modality-enhanced methods do not distinguish the distinct ways of ID and modality in reflecting user actions. By entangling these two causes, thus, these methods are unable to capture underlying rationales of user behaviors, leading to their inferior performance. Additionally, it serves as strong support for our motivation that it is imperative to disentangle ID and modality effects to improve SBR.  

Finally, the proposed \baby consistently outperforms all baseline methods in terms of all metrics in all datasets, which demonstrates its effectiveness for handling SBR. In particular, \baby achieves significant improvements over the best baselines in terms of Prec@20 and MRR@20 by 55.12\% and 18.01\% on Cellphones, 13.54\% and 2.27\% on Grocery, 22.75\% and 3.35\% on Sports, as well as 11.68\% and 19.65\% on Instacart. We believe that the overwhelming superiority of \baby over current start-of-the-art methods comes from its disentanglement for co-occurrence patterns of ID and fine-grained preferences of modality. Benefiting from distinguishing distinct effects of ID and modality, \baby is able to identify the specific rationale behind user behaviors, contributing to providing accurate predictions accordingly. 

\subsection{Effect of ID and modality representation learning (RQ2)}\label{sec:emb}

To effectively represent item ID and modality, we devise a co-occurrence representation schema where the co-occurrence patterns are explicitly incorporated into ID embeddings, and conduct modality alignment for unified modality representation via converting images into text. \babyx-co removes co-occurrence representation schema from \baby, that is, it implicitly encodes item co-occurrence patterns via neural networks (attention mechanism in this instance) as common paradigm~\cite{NARM, SR-GNN, BERT4Rec}. \babyx$_{cas}$ simply cascades embeddings of images and text to merge distinct modalities as in~\cite{lai@SIGIR2022, Pan@CIKM2022, MMMLP}. The performance of \baby and its variants is presented in Table~\ref{representation}, where Instacart is omitted due to its absence of image modality. 

It can be seen from Table~\ref{representation} that \baby outperforms both \babyx-co and \babyx$_{cas}$, which verifies the effectiveness of the devised techniques. We contend that explicitly learning paradigm enables \baby to adequately integrate co-occurrence patterns into item ID embeddings. Besides, converting distinct modalities into the same space and obtaining unified embeddings instead of representing them separately and fusing them at embedding level contributes to handling the semantic gap in different modalities. In addition, explicitly capturing co-occurrence patterns can bring large improvements for SBR, \ie~\babyx-co substantially underperforms \baby. It is reasonable since ID co-occurrence patterns are of great significance in modeling user actions and have dominated the field of recommender systems for over a decade~\cite{Yuan@SIGIR2023}.

\subsection{Effect of multi-view self-supervised disentanglement (RQ2)}\label{sec:multidisen}

\begin{figure}[t]
  \centering
  \includegraphics[width=0.98\linewidth]{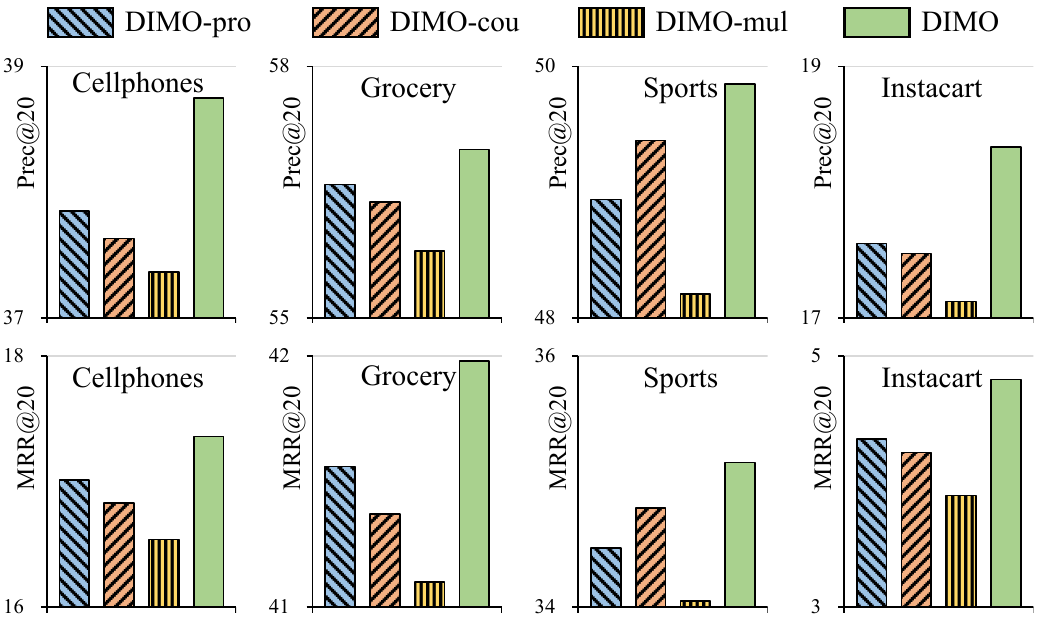}
  \caption{Effect of multi-view self-supervised disentanglement.}\label{mulitdisen}
\end{figure}

\begin{figure*}[ht]
  \centering
  \includegraphics[width=0.99\linewidth]{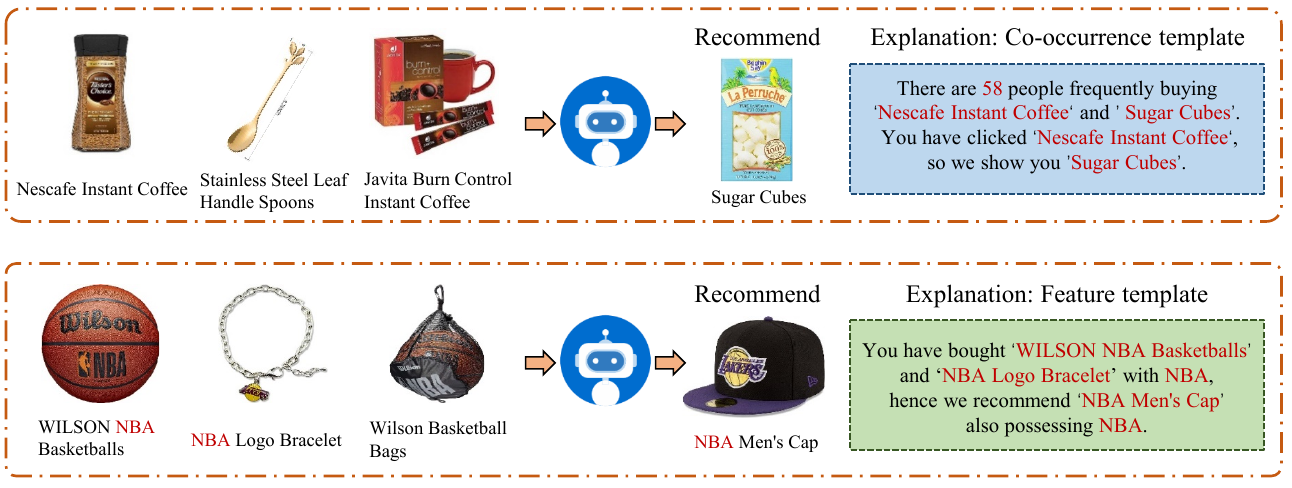}
  \caption{Case study for explainable session-based recommendation.}\label{casestudy}
\end{figure*}

\begin{figure}[t]
  \centering
  \includegraphics[width=0.99\linewidth]{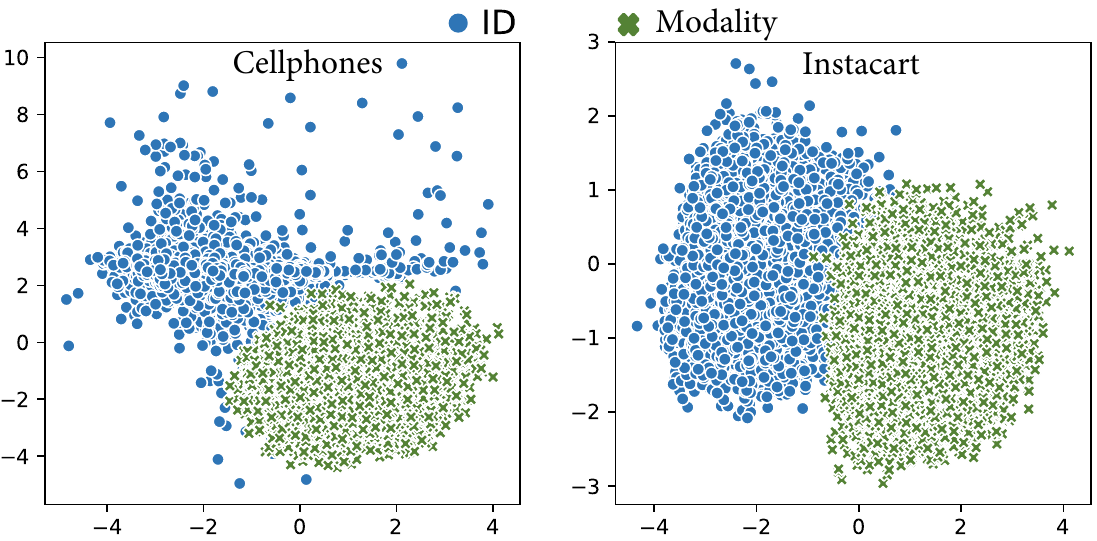}
  \caption{Visualization for item ID and modality embeddings of \baby. ID embeddings are represented by dots (blue), and modality embeddings are represented by cross (green).}\label{diseneffect}
\end{figure}

A multi-view self-supervised disentanglement, including proxy mechanism and counterfactual inference, is presented for distinguishing ID and modality effects in the absence of supervised signals at session level. To verify its effectiveness, the following variants of \baby are designed: \babyx-pro and \babyx-cou remove the proxy mechanism and counterfactual inference from \baby, respectively; \babyx-mul discards multi-view self-supervised disentanglement from \baby. 

The performance of above variants on all datasets is shown in~\fig~\ref{mulitdisen}, where the following insights can be obtained: (1) \baby achieves much better results than \babyx-mul, which indicates the effectiveness of the proposed multi-view self-supervised disentanglement on distinguishing ID and modality effects for SBR. We believe that ID and modality disentanglement enables \baby to comprehend specific causes behind user actions, contributing to revealing user intents; (2) Both of \babyx-pro and \babyx-cou are defeated by \baby, which proves the positive utility of the proxy mechanism and counterfactual inference in ID and modality disentanglement; (3) \babyx-pro and \babyx-cou exhibit performance gaps under different contexts, \eg~\babyx-pro outperforms \babyx-cou in Cellphones while doing worse in Sports. It signifies that the proxy mechanism and counterfactual inference are suitable for distinct situations. Thus, it is rational for \baby to employ these techniques together for effective ID and modality disentanglement.

\subsection{Disentanglement analysis (RQ3)}\label{sec:disen}

In order to intuitively examine the disentanglement effect of \baby, we plot item ID and modality embeddings of \baby in representative Cellphones and Instacart datasets. The PCA algorithm is employed to map embeddings into two-dimensional space for visualization. Due to space limits, we omit the results on Grocery and Sports, where we also obtained similar observations.  

As illustrated in~\fig~\ref{diseneffect}, ID and modality embeddings are far away from each other and separated into different regions. It indicates that our \baby achieves disentanglement for ID and modality. By distinguishing ID and modality at item and session levels, the proposed \baby is able to distinguish co-occurrence patterns of ID and fine-grained preferences of modality. Moreover, modality embeddings are concentrated, while ID embeddings are dispersed, as presented in Cellphones of~\fig~\ref{diseneffect}. Since item IDs do not carry item specific semantics, its embedding distribution primarily relies on data distribution in a dataset, where user behavior data usually present long-tail characteristic~\cite{Yang@SIGIR2023, Du@SIGIR2023}. In contrast, item modality records item semantic information, where items tend to possess similar attributes in a certain dataset. We believe that this is the reason for the distribution diversity on ID and modality embeddings. Also, this suggests the difference between co-occurrence patterns of ID and fine-grained preferences of modality in recommendation scenarios, which supports our motivation for ID and modality disentanglement once again.

\subsection{Case study for explainable SBR (RQ4)}\label{sec:explanation}

\begin{figure*}[ht]
  \centering
    \includegraphics[width=0.98\linewidth]{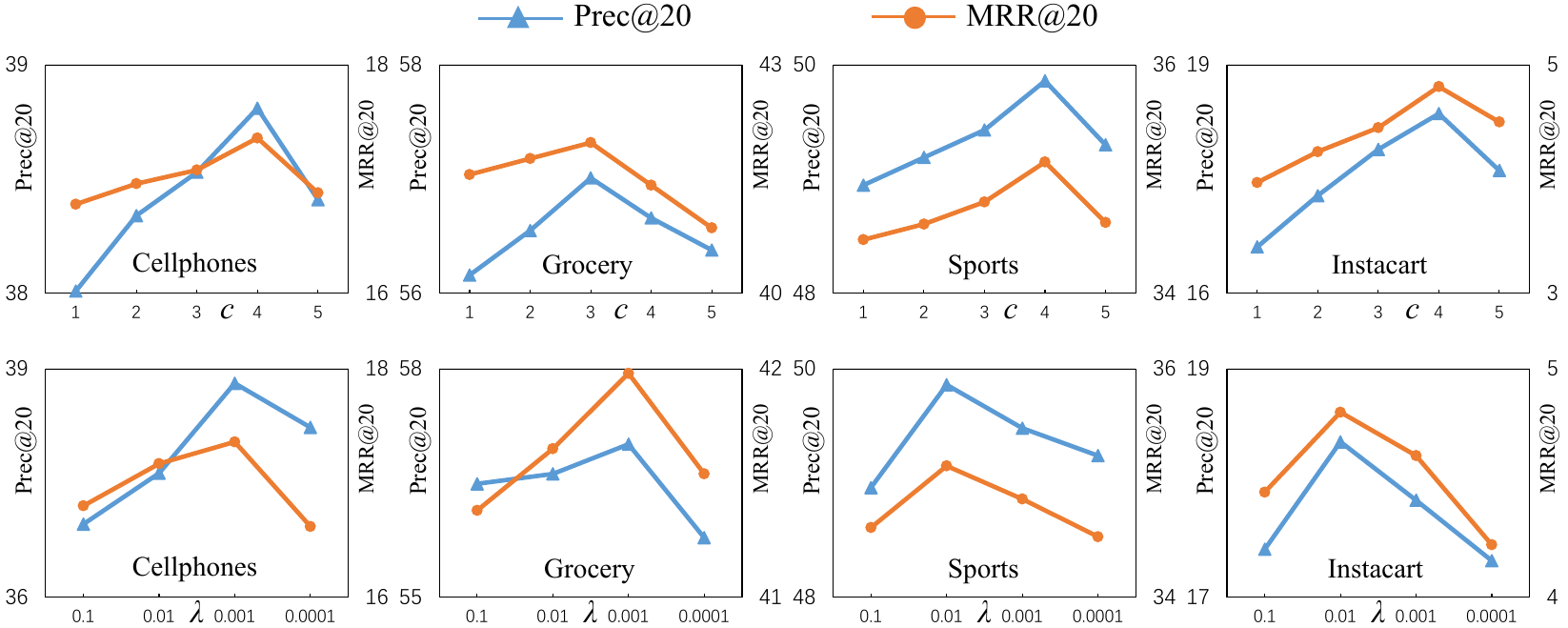}
  \caption{Impact of hyperparameters.}\label{hyperparameter}
\end{figure*}

In this part, we examine the interpretability of \baby in SBR via conducting case study where two instances are selected from Amazon datasets. For each case, we present items interacted by a user within a session, the items recommended by \baby, and the personalized explanations generated by \baby. With disentangled ID and modality causes, \baby can generate explanations from two perspectives, \ie co-occurrence patterns and fine-grained preferences. 

As presented in the first row of~\fig~\ref{casestudy}, \baby employs the co-occurrence template to generate explanations given that `coffee' and `sugar' are frequently purchased together. It is intuitive since user behaviors tend to possess conformity~\cite{Zheng@WWW2021} and co-occurrence associations are easy to understand~\cite{Wu@ICDE2023, Zheng@SIGIR2022}. 
Besides, \baby can also provide explanations using the feature template where user fine-grained preferences on certain features are captured. In the second row of~\fig~\ref{casestudy}, \baby finds that the user may be interested in NBA. Hence, it recommends a cap with NBA logo and provides explanations accordingly. Obviously, this greatly increases the serendipity of recommendations, which is of significance in real-life scenarios~\cite{Yang@SIGIR2023}. Overall, based on different recommendation rationales, explanations produced by \baby are meaningful and convincing, significantly contributing to improving user satisfaction.

\subsection{Hyperparameter Study (RQ5)}\label{sec:hyper}

As shown in~\fig~\ref{hyperparameter}, we plot the performance trends of the proposed \baby regarding two primary hyperparameters: (1) iterative times of global co-occurrence graph $c$, which determines the extent to which co-occurrence information is integrated into item ID embeddings; and (2) the coefficient of self-supervised tasks $\lambda$ in the loss function, which balances the recommendation task and auxiliary tasks. 

As illustrated in the first row of~\fig~\ref{hyperparameter}, we observe that with an increase in the value of $c$, \baby's performance initially improves but subsequently declines.  This trend can be attributed to the fact that higher values of $c$ enable \baby to integrate a greater amount of co-occurrence patterns into the item ID embeddings. However, excessive iterations may lead to an over-smoothing issue, where the embeddings lose their distinctive features, thereby negatively impacting the model's performance

As presented in the second row of~\fig~\ref{hyperparameter}, we empirically set $\lambda$ = 0.001 for Cellphones and Grocery, and $\lambda$ = 0.01 for Sports as well as Instacart. We contend that numerical scale and semantics of loss function in recommendation task and auxiliary tasks are different. For instance, recommendation task employs cross-entropy to highlight label items, while auxiliary tasks apply cosine similarity to drive similar/dissimilar embeddings close/apart. 
Given that various datasets possess different characteristics, they opt for distinct optimal values of $\lambda$.

\section{Conclusion and Future Work}
In this work, we identify two different rationales for modeling user behaviors, including co-occurrence patterns of ID and fine-grained preferences of modality (\eg text and images). 
Based on this observation, we propose a novel \baby that disentangles ID and modality effects to improve both accuracy and interpretability of session-based recommendation. To be specific, a co-occurrence representation schema is devised to explicitly incorporate co-occurrence patterns into ID embeddings. At the same time, we align different modalities via converting them into same semantic space for unified item modality representation. Afterward, a multi-view self-supervised disentanglement, including proxy mechanism and counterfactual inference, is presented to distinguish ID and modality effects within sessions in the absence of supervised signals. Based on disentangled causes, \baby forecasts users' next actions via causal inference. In addition, two kinds of templates, \ie co-occurrence template and feature template, are created to generate explanations. Extensive experiments on four real-world datasets demonstrate the overwhelming superiority of \baby over current state-of-the-art methods. Further analysis also indicates that the proposed \baby can generate meaningful explanations.  

Towards future work, we aim to apply our insights regarding the distinct effects of IDs and modalities to guide the development of Large Language Models (LLMs) in the realm of recommendation. 
A promising direction would be to integrate item ID co-occurrence patterns into LLMs to enhance the robustness and effectiveness of LLM-based recommendations.

\section*{Acknowledgement}
This work has been supported by the Natural Science Foundation of China (No.62076046, No.62376051, No.62076051). We would like to thank the anonymous reviewers for their valuable comments.

\newpage

\onecolumn
\begin{multicols}{2}
   \bibliographystyle{ACM-Reference-Format}
   \bibliography{main}


\begin{thebibliography}{48}


\ifx \showCODEN    \undefined \def \showCODEN     #1{\unskip}     \fi
\ifx \showDOI      \undefined \def \showDOI       #1{#1}\fi
\ifx \showISBNx    \undefined \def \showISBNx     #1{\unskip}     \fi
\ifx \showISBNxiii \undefined \def \showISBNxiii  #1{\unskip}     \fi
\ifx \showISSN     \undefined \def \showISSN      #1{\unskip}     \fi
\ifx \showLCCN     \undefined \def \showLCCN      #1{\unskip}     \fi
\ifx \shownote     \undefined \def \shownote      #1{#1}          \fi
\ifx \showarticletitle \undefined \def \showarticletitle #1{#1}   \fi
\ifx \showURL      \undefined \def \showURL       {\relax}        \fi
\providecommand\bibfield[2]{#2}
\providecommand\bibinfo[2]{#2}
\providecommand\natexlab[1]{#1}
\providecommand\showeprint[2][]{arXiv:#2}

\bibitem[Cai et~al\mbox{.}(2021)]%
        {cai@SIGIR2021}
\bibfield{author}{\bibinfo{person}{Renqin Cai}, \bibinfo{person}{Jibang Wu}, \bibinfo{person}{Aidan San}, \bibinfo{person}{Chong Wang}, {and} \bibinfo{person}{Hongning Wang}.} \bibinfo{year}{2021}\natexlab{}.
\newblock \showarticletitle{Category-aware Collaborative Sequential Recommendation}. In \bibinfo{booktitle}{\emph{{SIGIR}}}. \bibinfo{pages}{388--397}.
\newblock


\bibitem[Chen et~al\mbox{.}(2021)]%
        {Chen@IJCNN2021}
\bibfield{author}{\bibinfo{person}{Jiayi Chen}, \bibinfo{person}{Wen Wu}, \bibinfo{person}{Wenxin Hu}, \bibinfo{person}{Wei Zheng}, {and} \bibinfo{person}{Liang He}.} \bibinfo{year}{2021}\natexlab{}.
\newblock \showarticletitle{{SSR:} Explainable Session-based Recommendation}. In \bibinfo{booktitle}{\emph{{IJCNN}}}. \bibinfo{pages}{1--8}.
\newblock


\bibitem[Chen et~al\mbox{.}(2023)]%
        {Chen@SIGIR2023}
\bibfield{author}{\bibinfo{person}{Qian Chen}, \bibinfo{person}{Zhiqiang Guo}, \bibinfo{person}{Jianjun Li}, {and} \bibinfo{person}{Guohui Li}.} \bibinfo{year}{2023}\natexlab{}.
\newblock \showarticletitle{Knowledge-enhanced Multi-View Graph Neural Networks for Session-based Recommendation}. In \bibinfo{booktitle}{\emph{{SIGIR}}}. \bibinfo{pages}{352--361}.
\newblock


\bibitem[Chen and Wong(2020)]%
        {LESSR}
\bibfield{author}{\bibinfo{person}{Tianwen Chen} {and} \bibinfo{person}{Raymond~Chi{-}Wing Wong}.} \bibinfo{year}{2020}\natexlab{}.
\newblock \showarticletitle{Handling Information Loss of Graph Neural Networks for Session-based Recommendation}. In \bibinfo{booktitle}{\emph{{KDD}}}. \bibinfo{pages}{1172--1180}.
\newblock


\bibitem[Devlin et~al\mbox{.}(2019)]%
        {BERT}
\bibfield{author}{\bibinfo{person}{Jacob Devlin}, \bibinfo{person}{Ming{-}Wei Chang}, \bibinfo{person}{Kenton Lee}, {and} \bibinfo{person}{Kristina Toutanova}.} \bibinfo{year}{2019}\natexlab{}.
\newblock \showarticletitle{{BERT:} Pre-training of Deep Bidirectional Transformers for Language Understanding}. In \bibinfo{booktitle}{\emph{{NAACL-HLT}}}. \bibinfo{pages}{4171--4186}.
\newblock


\bibitem[Du et~al\mbox{.}(2023)]%
        {Du@SIGIR2023}
\bibfield{author}{\bibinfo{person}{Xinyu Du}, \bibinfo{person}{Huanhuan Yuan}, \bibinfo{person}{Pengpeng Zhao}, \bibinfo{person}{Jianfeng Qu}, \bibinfo{person}{Fuzhen Zhuang}, \bibinfo{person}{Guanfeng Liu}, \bibinfo{person}{Yanchi Liu}, {and} \bibinfo{person}{Victor~S. Sheng}.} \bibinfo{year}{2023}\natexlab{}.
\newblock \showarticletitle{Frequency Enhanced Hybrid Attention Network for Sequential Recommendation}. In \bibinfo{booktitle}{\emph{{SIGIR}}}. \bibinfo{pages}{78--88}.
\newblock


\bibitem[Geng et~al\mbox{.}(2022)]%
        {Geng@WWW2022}
\bibfield{author}{\bibinfo{person}{Shijie Geng}, \bibinfo{person}{Zuohui Fu}, \bibinfo{person}{Juntao Tan}, \bibinfo{person}{Yingqiang Ge}, \bibinfo{person}{Gerard de Melo}, {and} \bibinfo{person}{Yongfeng Zhang}.} \bibinfo{year}{2022}\natexlab{}.
\newblock \showarticletitle{Path Language Modeling over Knowledge Graphs for Explainable Recommendation}. In \bibinfo{booktitle}{\emph{{WWW}}}. \bibinfo{pages}{946--955}.
\newblock


\bibitem[Guo et~al\mbox{.}(2022)]%
        {guo@WSDM2022}
\bibfield{author}{\bibinfo{person}{Jiayan Guo}, \bibinfo{person}{Yaming Yang}, \bibinfo{person}{Xiangchen Song}, \bibinfo{person}{Yuan Zhang}, \bibinfo{person}{Yujing Wang}, \bibinfo{person}{Jing Bai}, {and} \bibinfo{person}{Yan Zhang}.} \bibinfo{year}{2022}\natexlab{}.
\newblock \showarticletitle{Learning Multi-granularity Consecutive User Intent Unit for Session-based Recommendation}. In \bibinfo{booktitle}{\emph{{WSDM}}}. \bibinfo{pages}{343--352}.
\newblock


\bibitem[Han et~al\mbox{.}(2022)]%
        {han@SIGIR2022}
\bibfield{author}{\bibinfo{person}{Qilong Han}, \bibinfo{person}{Chi Zhang}, \bibinfo{person}{Rui Chen}, \bibinfo{person}{Riwei Lai}, \bibinfo{person}{Hongtao Song}, {and} \bibinfo{person}{Li Li}.} \bibinfo{year}{2022}\natexlab{}.
\newblock \showarticletitle{Multi-Faceted Global Item Relation Learning for Session-Based Recommendation}. In \bibinfo{booktitle}{\emph{{SIGIR}}}. \bibinfo{pages}{1705--1715}.
\newblock


\bibitem[Hidasi et~al\mbox{.}(2016)]%
        {GRU4Rec}
\bibfield{author}{\bibinfo{person}{Bal{\'{a}}zs Hidasi}, \bibinfo{person}{Alexandros Karatzoglou}, \bibinfo{person}{Linas Baltrunas}, {and} \bibinfo{person}{Domonkos Tikk}.} \bibinfo{year}{2016}\natexlab{}.
\newblock \showarticletitle{Session-based Recommendations with Recurrent Neural Networks}. In \bibinfo{booktitle}{\emph{{ICLR}}}.
\newblock


\bibitem[Hou et~al\mbox{.}(2022)]%
        {hou@KDD2022}
\bibfield{author}{\bibinfo{person}{Yupeng Hou}, \bibinfo{person}{Shanlei Mu}, \bibinfo{person}{Wayne~Xin Zhao}, \bibinfo{person}{Yaliang Li}, \bibinfo{person}{Bolin Ding}, {and} \bibinfo{person}{Ji{-}Rong Wen}.} \bibinfo{year}{2022}\natexlab{}.
\newblock \showarticletitle{Towards Universal Sequence Representation Learning for Recommender Systems}. In \bibinfo{booktitle}{\emph{{KDD}}}. \bibinfo{pages}{585--593}.
\newblock


\bibitem[Hu et~al\mbox{.}(2023)]%
        {Hu@CIKM2023}
\bibfield{author}{\bibinfo{person}{Hengchang Hu}, \bibinfo{person}{Wei Guo}, \bibinfo{person}{Yong Liu}, {and} \bibinfo{person}{Min{-}Yen Kan}.} \bibinfo{year}{2023}\natexlab{}.
\newblock \showarticletitle{Adaptive Multi-Modalities Fusion in Sequential Recommendation Systems}. In \bibinfo{booktitle}{\emph{{CIKM}}}. \bibinfo{publisher}{{ACM}}, \bibinfo{pages}{843--853}.
\newblock


\bibitem[Lai et~al\mbox{.}(2022)]%
        {lai@SIGIR2022}
\bibfield{author}{\bibinfo{person}{Siqi Lai}, \bibinfo{person}{Erli Meng}, \bibinfo{person}{Fan Zhang}, \bibinfo{person}{Chenliang Li}, \bibinfo{person}{Bin Wang}, {and} \bibinfo{person}{Aixin Sun}.} \bibinfo{year}{2022}\natexlab{}.
\newblock \showarticletitle{An Attribute-Driven Mirror Graph Network for Session-based Recommendation}. In \bibinfo{booktitle}{\emph{{SIGIR}}}. \bibinfo{pages}{1674--1683}.
\newblock


\bibitem[Li et~al\mbox{.}(2017)]%
        {NARM}
\bibfield{author}{\bibinfo{person}{Jing Li}, \bibinfo{person}{Pengjie Ren}, \bibinfo{person}{Zhumin Chen}, \bibinfo{person}{Zhaochun Ren}, \bibinfo{person}{Tao Lian}, {and} \bibinfo{person}{Jun Ma}.} \bibinfo{year}{2017}\natexlab{}.
\newblock \showarticletitle{Neural Attentive Session-based Recommendation}. In \bibinfo{booktitle}{\emph{{CIKM}}}. \bibinfo{pages}{1419--1428}.
\newblock


\bibitem[Li et~al\mbox{.}(2023a)]%
        {Li@KDD2023}
\bibfield{author}{\bibinfo{person}{Jiacheng Li}, \bibinfo{person}{Ming Wang}, \bibinfo{person}{Jin Li}, \bibinfo{person}{Jinmiao Fu}, \bibinfo{person}{Xin Shen}, \bibinfo{person}{Jingbo Shang}, {and} \bibinfo{person}{Julian~J. McAuley}.} \bibinfo{year}{2023}\natexlab{a}.
\newblock \showarticletitle{Text Is All You Need: Learning Language Representations for Sequential Recommendation}. In \bibinfo{booktitle}{\emph{{KDD}}}. \bibinfo{pages}{1258--1267}.
\newblock


\bibitem[Li et~al\mbox{.}(2023b)]%
        {Li@WSDM2023}
\bibfield{author}{\bibinfo{person}{Zihao Li}, \bibinfo{person}{Xianzhi Wang}, \bibinfo{person}{Chao Yang}, \bibinfo{person}{Lina Yao}, \bibinfo{person}{Julian~J. McAuley}, {and} \bibinfo{person}{Guandong Xu}.} \bibinfo{year}{2023}\natexlab{b}.
\newblock \showarticletitle{Exploiting Explicit and Implicit Item relationships for Session-based Recommendation}. In \bibinfo{booktitle}{\emph{{WSDM}}}. \bibinfo{pages}{553--561}.
\newblock


\bibitem[Liang et~al\mbox{.}(2023)]%
        {MMMLP}
\bibfield{author}{\bibinfo{person}{Jiahao Liang}, \bibinfo{person}{Xiangyu Zhao}, \bibinfo{person}{Muyang Li}, \bibinfo{person}{Zijian Zhang}, \bibinfo{person}{Wanyu Wang}, \bibinfo{person}{Haochen Liu}, {and} \bibinfo{person}{Zitao Liu}.} \bibinfo{year}{2023}\natexlab{}.
\newblock \showarticletitle{{MMMLP:} Multi-modal Multilayer Perceptron for Sequential Recommendations}. In \bibinfo{booktitle}{\emph{{WWW}}}. \bibinfo{pages}{1109--1117}.
\newblock


\bibitem[Liu et~al\mbox{.}(2018)]%
        {STAMP}
\bibfield{author}{\bibinfo{person}{Qiao Liu}, \bibinfo{person}{Yifu Zeng}, \bibinfo{person}{Refuoe Mokhosi}, {and} \bibinfo{person}{Haibin Zhang}.} \bibinfo{year}{2018}\natexlab{}.
\newblock \showarticletitle{{STAMP:} Short-Term Attention/Memory Priority Model for Session-based Recommendation}. In \bibinfo{booktitle}{\emph{{KDD}}}. \bibinfo{pages}{1831--1839}.
\newblock


\bibitem[Liu et~al\mbox{.}(2023)]%
        {Liu@CIKM2023Text}
\bibfield{author}{\bibinfo{person}{Zhenghao Liu}, \bibinfo{person}{Sen Mei}, \bibinfo{person}{Chenyan Xiong}, \bibinfo{person}{Xiaohua Li}, \bibinfo{person}{Shi Yu}, \bibinfo{person}{Zhiyuan Liu}, \bibinfo{person}{Yu Gu}, {and} \bibinfo{person}{Ge Yu}.} \bibinfo{year}{2023}\natexlab{}.
\newblock \showarticletitle{Text Matching Improves Sequential Recommendation by Reducing Popularity Biases}. In \bibinfo{booktitle}{\emph{{CIKM}}}. \bibinfo{publisher}{{ACM}}, \bibinfo{pages}{1534--1544}.
\newblock


\bibitem[Ma et~al\mbox{.}(2019)]%
        {Ma@NIPS2019}
\bibfield{author}{\bibinfo{person}{Jianxin Ma}, \bibinfo{person}{Chang Zhou}, \bibinfo{person}{Peng Cui}, \bibinfo{person}{Hongxia Yang}, {and} \bibinfo{person}{Wenwu Zhu}.} \bibinfo{year}{2019}\natexlab{}.
\newblock \showarticletitle{Learning Disentangled Representations for Recommendation}. In \bibinfo{booktitle}{\emph{NeurIPS}}. \bibinfo{pages}{5712--5723}.
\newblock


\bibitem[Pan et~al\mbox{.}(2022)]%
        {Pan@CIKM2022}
\bibfield{author}{\bibinfo{person}{Xingyu Pan}, \bibinfo{person}{Yushuo Chen}, \bibinfo{person}{Changxin Tian}, \bibinfo{person}{Zihan Lin}, \bibinfo{person}{Jinpeng Wang}, \bibinfo{person}{He Hu}, {and} \bibinfo{person}{Wayne~Xin Zhao}.} \bibinfo{year}{2022}\natexlab{}.
\newblock \showarticletitle{Multimodal Meta-Learning for Cold-Start Sequential Recommendation}. In \bibinfo{booktitle}{\emph{{CIKM}}}. \bibinfo{pages}{3421--3430}.
\newblock


\bibitem[Pang et~al\mbox{.}(2022)]%
        {pang@WSDM2022}
\bibfield{author}{\bibinfo{person}{Yitong Pang}, \bibinfo{person}{Lingfei Wu}, \bibinfo{person}{Qi Shen}, \bibinfo{person}{Yiming Zhang}, \bibinfo{person}{Zhihua Wei}, \bibinfo{person}{Fangli Xu}, \bibinfo{person}{Ethan Chang}, \bibinfo{person}{Bo Long}, {and} \bibinfo{person}{Jian Pei}.} \bibinfo{year}{2022}\natexlab{}.
\newblock \showarticletitle{Heterogeneous Global Graph Neural Networks for Personalized Session-based Recommendation}. In \bibinfo{booktitle}{\emph{{WSDM}}}. \bibinfo{pages}{775--783}.
\newblock


\bibitem[Peintner et~al\mbox{.}(2023)]%
        {Andreas@RecSys2023}
\bibfield{author}{\bibinfo{person}{Andreas Peintner}, \bibinfo{person}{Amir~Reza Mohammadi}, {and} \bibinfo{person}{Eva Zangerle}.} \bibinfo{year}{2023}\natexlab{}.
\newblock \showarticletitle{{SPARE:} Shortest Path Global Item Relations for Efficient Session-based Recommendation}. In \bibinfo{booktitle}{\emph{RecSys}}. \bibinfo{pages}{58--69}.
\newblock


\bibitem[Qiao et~al\mbox{.}(2023)]%
        {Qiao@CIKM2023}
\bibfield{author}{\bibinfo{person}{Shutong Qiao}, \bibinfo{person}{Wei Zhou}, \bibinfo{person}{Junhao Wen}, \bibinfo{person}{Hongyu Zhang}, {and} \bibinfo{person}{Min Gao}.} \bibinfo{year}{2023}\natexlab{}.
\newblock \showarticletitle{Bi-channel Multiple Sparse Graph Attention Networks for Session-based Recommendation}. In \bibinfo{booktitle}{\emph{{CIKM}}}. \bibinfo{publisher}{{ACM}}, \bibinfo{pages}{2075--2084}.
\newblock


\bibitem[Song et~al\mbox{.}(2023)]%
        {Song@WWW2023}
\bibfield{author}{\bibinfo{person}{Wenzhuo Song}, \bibinfo{person}{Shoujin Wang}, \bibinfo{person}{Yan Wang}, \bibinfo{person}{Kunpeng Liu}, \bibinfo{person}{Xueyan Liu}, {and} \bibinfo{person}{Minghao Yin}.} \bibinfo{year}{2023}\natexlab{}.
\newblock \showarticletitle{A Counterfactual Collaborative Session-based Recommender System}. In \bibinfo{booktitle}{\emph{{WWW}}}. \bibinfo{pages}{971--982}.
\newblock


\bibitem[Su et~al\mbox{.}(2023)]%
        {Su@WWW2023}
\bibfield{author}{\bibinfo{person}{Jiajie Su}, \bibinfo{person}{Chaochao Chen}, \bibinfo{person}{Weiming Liu}, \bibinfo{person}{Fei Wu}, \bibinfo{person}{Xiaolin Zheng}, {and} \bibinfo{person}{Haoming Lyu}.} \bibinfo{year}{2023}\natexlab{}.
\newblock \showarticletitle{Enhancing Hierarchy-Aware Graph Networks with Deep Dual Clustering for Session-based Recommendation}. In \bibinfo{booktitle}{\emph{{WWW}}}. \bibinfo{pages}{165--176}.
\newblock


\bibitem[Sun et~al\mbox{.}(2019)]%
        {BERT4Rec}
\bibfield{author}{\bibinfo{person}{Fei Sun}, \bibinfo{person}{Jun Liu}, \bibinfo{person}{Jian Wu}, \bibinfo{person}{Changhua Pei}, \bibinfo{person}{Xiao Lin}, \bibinfo{person}{Wenwu Ou}, {and} \bibinfo{person}{Peng Jiang}.} \bibinfo{year}{2019}\natexlab{}.
\newblock \showarticletitle{BERT4Rec: Sequential Recommendation with Bidirectional Encoder Representations from Transformer}. In \bibinfo{booktitle}{\emph{{CIKM}}}. \bibinfo{pages}{1441--1450}.
\newblock


\bibitem[Wang et~al\mbox{.}(2022)]%
        {Wang@CS2022}
\bibfield{author}{\bibinfo{person}{Shoujin Wang}, \bibinfo{person}{Longbing Cao}, \bibinfo{person}{Yan Wang}, \bibinfo{person}{Quan~Z. Sheng}, \bibinfo{person}{Mehmet~A. Orgun}, {and} \bibinfo{person}{Defu Lian}.} \bibinfo{year}{2022}\natexlab{}.
\newblock \showarticletitle{A Survey on Session-based Recommender Systems}.
\newblock \bibinfo{journal}{\emph{{ACM} Comput. Surv.}} (\bibinfo{year}{2022}), \bibinfo{pages}{154:1--154:38}.
\newblock


\bibitem[Wang et~al\mbox{.}(2020)]%
        {GCE-GNN}
\bibfield{author}{\bibinfo{person}{Ziyang Wang}, \bibinfo{person}{Wei Wei}, \bibinfo{person}{Gao Cong}, \bibinfo{person}{Xiao{-}Li Li}, \bibinfo{person}{Xianling Mao}, {and} \bibinfo{person}{Minghui Qiu}.} \bibinfo{year}{2020}\natexlab{}.
\newblock \showarticletitle{Global Context Enhanced Graph Neural Networks for Session-based Recommendation}. In \bibinfo{booktitle}{\emph{{SIGIR}}}. \bibinfo{pages}{169--178}.
\newblock


\bibitem[Wu et~al\mbox{.}(2023)]%
        {Wu@ICDE2023}
\bibfield{author}{\bibinfo{person}{Huizi Wu}, \bibinfo{person}{Hui Fang}, \bibinfo{person}{Zhu Sun}, \bibinfo{person}{Cong Geng}, \bibinfo{person}{Xinyu Kong}, {and} \bibinfo{person}{Yew{-}Soon Ong}.} \bibinfo{year}{2023}\natexlab{}.
\newblock \showarticletitle{A Generic Reinforced Explainable Framework with Knowledge Graph for Session-based Recommendation}. In \bibinfo{booktitle}{\emph{{ICDE}}}. \bibinfo{pages}{1260--1272}.
\newblock


\bibitem[Wu et~al\mbox{.}(2019)]%
        {SR-GNN}
\bibfield{author}{\bibinfo{person}{Shu Wu}, \bibinfo{person}{Yuyuan Tang}, \bibinfo{person}{Yanqiao Zhu}, \bibinfo{person}{Liang Wang}, \bibinfo{person}{Xing Xie}, {and} \bibinfo{person}{Tieniu Tan}.} \bibinfo{year}{2019}\natexlab{}.
\newblock \showarticletitle{Session-Based Recommendation with Graph Neural Networks}. In \bibinfo{booktitle}{\emph{{AAAI}}}. \bibinfo{pages}{346--353}.
\newblock


\bibitem[Yang et~al\mbox{.}(2023)]%
        {Yang@SIGIR2023}
\bibfield{author}{\bibinfo{person}{Heeyoon Yang}, \bibinfo{person}{YunSeok Choi}, \bibinfo{person}{Gahyung Kim}, {and} \bibinfo{person}{Jee{-}Hyong Lee}.} \bibinfo{year}{2023}\natexlab{}.
\newblock \showarticletitle{{LOAM:} Improving Long-tail Session-based Recommendation via Niche Walk Augmentation and Tail Session Mixup}. In \bibinfo{booktitle}{\emph{{SIGIR}}}. \bibinfo{pages}{527--536}.
\newblock


\bibitem[Yuan et~al\mbox{.}(2023)]%
        {Yuan@SIGIR2023}
\bibfield{author}{\bibinfo{person}{Zheng Yuan}, \bibinfo{person}{Fajie Yuan}, \bibinfo{person}{Yu Song}, \bibinfo{person}{Youhua Li}, \bibinfo{person}{Junchen Fu}, \bibinfo{person}{Fei Yang}, \bibinfo{person}{Yunzhu Pan}, {and} \bibinfo{person}{Yongxin Ni}.} \bibinfo{year}{2023}\natexlab{}.
\newblock \showarticletitle{Where to Go Next for Recommender Systems? {ID-} vs. Modality-based Recommender Models Revisited}. In \bibinfo{booktitle}{\emph{{SIGIR}}}. \bibinfo{pages}{2639--2649}.
\newblock


\bibitem[Zhang et~al\mbox{.}(2023a)]%
        {Zhang@WSDM2023}
\bibfield{author}{\bibinfo{person}{Peiyan Zhang}, \bibinfo{person}{Jiayan Guo}, \bibinfo{person}{Chaozhuo Li}, \bibinfo{person}{Yueqi Xie}, \bibinfo{person}{Jaeboum Kim}, \bibinfo{person}{Yan Zhang}, \bibinfo{person}{Xing Xie}, \bibinfo{person}{Haohan Wang}, {and} \bibinfo{person}{Sunghun Kim}.} \bibinfo{year}{2023}\natexlab{a}.
\newblock \showarticletitle{Efficiently Leveraging Multi-level User Intent for Session-based Recommendation via Atten-Mixer Network}. In \bibinfo{booktitle}{\emph{{WSDM}}}. \bibinfo{pages}{168--176}.
\newblock


\bibitem[Zhang et~al\mbox{.}(2022a)]%
        {DIDN}
\bibfield{author}{\bibinfo{person}{Xiaokun Zhang}, \bibinfo{person}{Hongfei Lin}, \bibinfo{person}{Bo Xu}, \bibinfo{person}{Chenliang Li}, \bibinfo{person}{Yuan Lin}, \bibinfo{person}{Haifeng Liu}, {and} \bibinfo{person}{Fenglong Ma}.} \bibinfo{year}{2022}\natexlab{a}.
\newblock \showarticletitle{Dynamic intent-aware iterative denoising network for session-based recommendation}.
\newblock \bibinfo{journal}{\emph{Inf. Process. Manag.}} \bibinfo{volume}{59}, \bibinfo{number}{3} (\bibinfo{year}{2022}), \bibinfo{pages}{102936}.
\newblock


\bibitem[Zhang et~al\mbox{.}(2023b)]%
        {Zhang@WSDM2023Disen}
\bibfield{author}{\bibinfo{person}{Xiaoying Zhang}, \bibinfo{person}{Hongning Wang}, {and} \bibinfo{person}{Hang Li}.} \bibinfo{year}{2023}\natexlab{b}.
\newblock \showarticletitle{Disentangled Representation for Diversified Recommendations}. In \bibinfo{booktitle}{\emph{{WSDM}}}. \bibinfo{pages}{490--498}.
\newblock


\bibitem[Zhang et~al\mbox{.}(2024a)]%
        {zhang2024side}
\bibfield{author}{\bibinfo{person}{Xiaokun Zhang}, \bibinfo{person}{Bo Xu}, \bibinfo{person}{Chenliang Li}, \bibinfo{person}{Yao Zhou}, \bibinfo{person}{Liangyue Li}, {and} \bibinfo{person}{Hongfei Lin}.} \bibinfo{year}{2024}\natexlab{a}.
\newblock \showarticletitle{Side Information-Driven Session-based Recommendation: A Survey}.
\newblock \bibinfo{journal}{\emph{arXiv preprint arXiv:2402.17129}} (\bibinfo{year}{2024}).
\newblock


\bibitem[Zhang et~al\mbox{.}(2023d)]%
        {BiPNet}
\bibfield{author}{\bibinfo{person}{Xiaokun Zhang}, \bibinfo{person}{Bo Xu}, \bibinfo{person}{Fenglong Ma}, \bibinfo{person}{Chenliang Li}, \bibinfo{person}{Yuan Lin}, {and} \bibinfo{person}{Hongfei Lin}.} \bibinfo{year}{2023}\natexlab{d}.
\newblock \showarticletitle{Bi-preference Learning Heterogeneous Hypergraph Networks for Session-based Recommendation}.
\newblock \bibinfo{journal}{\emph{ACM Trans. Inf. Syst.}} \bibinfo{volume}{42}, \bibinfo{number}{3}, Article \bibinfo{articleno}{68} (\bibinfo{year}{2023}), \bibinfo{numpages}{28}~pages.
\newblock


\bibitem[Zhang et~al\mbox{.}(2024b)]%
        {MMSBR}
\bibfield{author}{\bibinfo{person}{Xiaokun Zhang}, \bibinfo{person}{Bo Xu}, \bibinfo{person}{Fenglong Ma}, \bibinfo{person}{Chenliang Li}, \bibinfo{person}{Liang Yang}, {and} \bibinfo{person}{Hongfei Lin}.} \bibinfo{year}{2024}\natexlab{b}.
\newblock \showarticletitle{Beyond Co-Occurrence: Multi-Modal Session-Based Recommendation}.
\newblock \bibinfo{journal}{\emph{{IEEE} Trans. Knowl. Data Eng.}} \bibinfo{volume}{36}, \bibinfo{number}{4} (\bibinfo{year}{2024}), \bibinfo{pages}{1450--1462}.
\newblock


\bibitem[Zhang et~al\mbox{.}(2022b)]%
        {CoHHN}
\bibfield{author}{\bibinfo{person}{Xiaokun Zhang}, \bibinfo{person}{Bo Xu}, \bibinfo{person}{Liang Yang}, \bibinfo{person}{Chenliang Li}, \bibinfo{person}{Fenglong Ma}, \bibinfo{person}{Haifeng Liu}, {and} \bibinfo{person}{Hongfei Lin}.} \bibinfo{year}{2022}\natexlab{b}.
\newblock \showarticletitle{Price {DOES} Matter!: Modeling Price and Interest Preferences in Session-based Recommendation}. In \bibinfo{booktitle}{\emph{{SIGIR}}}. \bibinfo{pages}{1684--1693}.
\newblock


\bibitem[Zhang et~al\mbox{.}(2014)]%
        {Zhang@SIGIR2014}
\bibfield{author}{\bibinfo{person}{Yongfeng Zhang}, \bibinfo{person}{Guokun Lai}, \bibinfo{person}{Min Zhang}, \bibinfo{person}{Yi Zhang}, \bibinfo{person}{Yiqun Liu}, {and} \bibinfo{person}{Shaoping Ma}.} \bibinfo{year}{2014}\natexlab{}.
\newblock \showarticletitle{Explicit factor models for explainable recommendation based on phrase-level sentiment analysis}. In \bibinfo{booktitle}{\emph{{SIGIR}}}. \bibinfo{pages}{83--92}.
\newblock


\bibitem[Zhang et~al\mbox{.}(2023c)]%
        {Zhang@KDD2023}
\bibfield{author}{\bibinfo{person}{Yipeng Zhang}, \bibinfo{person}{Xin Wang}, \bibinfo{person}{Hong Chen}, {and} \bibinfo{person}{Wenwu Zhu}.} \bibinfo{year}{2023}\natexlab{c}.
\newblock \showarticletitle{Adaptive Disentangled Transformer for Sequential Recommendation}. In \bibinfo{booktitle}{\emph{{KDD}}}. \bibinfo{pages}{3434--3445}.
\newblock


\bibitem[Zheng et~al\mbox{.}(2022b)]%
        {Zheng@SIGIR2022}
\bibfield{author}{\bibinfo{person}{Jiayin Zheng}, \bibinfo{person}{Juanyun Mai}, {and} \bibinfo{person}{Yanlong Wen}.} \bibinfo{year}{2022}\natexlab{b}.
\newblock \showarticletitle{Explainable Session-based Recommendation with Meta-path Guided Instances and Self-Attention Mechanism}. In \bibinfo{booktitle}{\emph{{SIGIR}}}. \bibinfo{pages}{2555--2559}.
\newblock


\bibitem[Zheng et~al\mbox{.}(2022a)]%
        {Zheng@WWW2022}
\bibfield{author}{\bibinfo{person}{Yu Zheng}, \bibinfo{person}{Chen Gao}, \bibinfo{person}{Jianxin Chang}, \bibinfo{person}{Yanan Niu}, \bibinfo{person}{Yang Song}, \bibinfo{person}{Depeng Jin}, {and} \bibinfo{person}{Yong Li}.} \bibinfo{year}{2022}\natexlab{a}.
\newblock \showarticletitle{Disentangling Long and Short-Term Interests for Recommendation}. In \bibinfo{booktitle}{\emph{{WWW}}}. \bibinfo{pages}{2256--2267}.
\newblock


\bibitem[Zheng et~al\mbox{.}(2021)]%
        {Zheng@WWW2021}
\bibfield{author}{\bibinfo{person}{Yu Zheng}, \bibinfo{person}{Chen Gao}, \bibinfo{person}{Xiang Li}, \bibinfo{person}{Xiangnan He}, \bibinfo{person}{Yong Li}, {and} \bibinfo{person}{Depeng Jin}.} \bibinfo{year}{2021}\natexlab{}.
\newblock \showarticletitle{Disentangling User Interest and Conformity for Recommendation with Causal Embedding}. In \bibinfo{booktitle}{\emph{{WWW}}}. \bibinfo{pages}{2980--2991}.
\newblock


\bibitem[Zhou et~al\mbox{.}(2020)]%
        {Zhou@CIKM2020}
\bibfield{author}{\bibinfo{person}{Kun Zhou}, \bibinfo{person}{Hui Wang}, \bibinfo{person}{Wayne~Xin Zhao}, \bibinfo{person}{Yutao Zhu}, \bibinfo{person}{Sirui Wang}, \bibinfo{person}{Fuzheng Zhang}, \bibinfo{person}{Zhongyuan Wang}, {and} \bibinfo{person}{Ji{-}Rong Wen}.} \bibinfo{year}{2020}\natexlab{}.
\newblock \showarticletitle{S3-Rec: Self-Supervised Learning for Sequential Recommendation with Mutual Information Maximization}. In \bibinfo{booktitle}{\emph{{CIKM}}}. \bibinfo{pages}{1893--1902}.
\newblock


\bibitem[Zhou et~al\mbox{.}(2022)]%
        {zhou@WWW2022}
\bibfield{author}{\bibinfo{person}{Kun Zhou}, \bibinfo{person}{Hui Yu}, \bibinfo{person}{Wayne~Xin Zhao}, {and} \bibinfo{person}{Ji{-}Rong Wen}.} \bibinfo{year}{2022}\natexlab{}.
\newblock \showarticletitle{Filter-enhanced {MLP} is All You Need for Sequential Recommendation}. In \bibinfo{booktitle}{\emph{{WWW}}}. \bibinfo{pages}{2388--2399}.
\newblock


\bibitem[Zhou et~al\mbox{.}(2023)]%
        {Zhou@CIKM2023}
\bibfield{author}{\bibinfo{person}{Peilin Zhou}, \bibinfo{person}{Qichen Ye}, \bibinfo{person}{Yueqi Xie}, \bibinfo{person}{Jingqi Gao}, \bibinfo{person}{Shoujin Wang}, \bibinfo{person}{Jae~Boum Kim}, \bibinfo{person}{Chenyu You}, {and} \bibinfo{person}{Sunghun Kim}.} \bibinfo{year}{2023}\natexlab{}.
\newblock \showarticletitle{Attention Calibration for Transformer-based Sequential Recommendation}. In \bibinfo{booktitle}{\emph{{CIKM}}}. \bibinfo{publisher}{{ACM}}, \bibinfo{pages}{3595--3605}.
\newblock


\end{thebibliography}
\end{multicols}

\end{document}